\documentclass[12pt,preprint]{aastex}
\usepackage{url,color}
\usepackage[]{natbib}

\begin{document}

\title{Cameras a Million Miles Apart: Stereoscopic Imaging Potential with the Hubble and James Webb Space Telescopes}

\author{
Joel D. Green\altaffilmark{1,2},
Johannes Burge\altaffilmark{3},
John A. Stansberry\altaffilmark{1},
\& Bonnie Meinke\altaffilmark{1}
}

\altaffiltext{1}{Space Telescope Science Institute, Baltimore, MD, USA}
\altaffiltext{2}{Dept of Astronomy, The University of Texas at Austin, Austin, TX 78712, USA}
\altaffiltext{3}{Dept of Psychology, Neuroscience Graduate Group, University of Pennsylvania, Philadelphia, PA}

\begin{abstract}

The two most powerful optical/IR telescopes in history -- NASA's Hubble and James Webb Space Telescopes -- will be in space at the same time.  We have a unique opportunity to leverage the 1.5 million kilometer separation between the two telescopic nodal points to obtain simultaneously captured stereoscopic images of asteroids, comets, moons and planets in our Solar System. Given the recent resurgence in stereo-3D movies and the recent emergence of VR-enabled mobile devices, these stereoscopic images provide a unique opportunity to engage the public with unprecedented views of various Solar System objects. Here, we present the technical requirements for acquiring stereoscopic images of Solar System objects, given the constraints of the telescopic equipment and the orbits of the target objects, and we present a handful of examples.

\end{abstract}

\section{Introduction}

When encountering astronomical imagery, the public can be significantly disconnected from the visualizations adopted by scientists. The experience of seeing Saturn and its ring system with a simple ``backyard'' telescope is thrilling for many viewers. But typical telescopic images lack binocular depth cues because each image is captured from a single viewing position. It is possible to achieve stereo views of a Solar System object with astronomical parallax (waiting 0.5 yr between images), but this does not allow for {\it simultaneous} imaging of a non-sidereal or transient event. One way to achieve a vivid impression of depth is to image target objects stereoscopically.

Stereopsis is the perception of depth from binocular disparity. Binocular disparities are the small, subtle differences between the two eyes' images due to the eyes' different vantage points on a 3D scene. Binocular disparity is the most precise depth cue available to the human visual system. Under optimal conditions, the human visual system is capable of detecting binocular disparities as small as 5 arcsec \citep{blakemore70,cormack91,burge14}, approximately 1/6 the width of an individual foveal photoreceptor \citep{curcio90}. Given a typical human interocular distance (65mm), this disparity detection threshold means that a human observer can reliably detect depth differences as small as $\sim$ 1/3 mm at a distance of 1m. The binocular disparity associated with a given depth difference decreases approximately with the square of the distance, and increases linearly with the physical separation between the imaging devices (e.g. eyes). Thus, the larger the physical separation between the imaging devices, the greater the potential for resolving small depth differences. 

Space telescopes have the potential to achieve enormous physical separations. In the next several years, the two highest resolution space telescopes in history--the Hubble Space Telescope (HST) and the soon-to-be-launched James Webb Space Telescope (JWST)--will be in space simultaneously. The physical separation between these telescopes will average 1.5 million kilometers, with a range from 1.3 to 1.7 million kilometers; and the spatial resolution of each telescope in the optical/near-infrared is $\sim$ 0.07 arcsec\footnote{With a 6.5m aperture, JWST is not diffraction limited at wavelengths $<$ 2 $\mu$m.}. The physical separation between the telescopes will be large enough, given the spatial resolution of the telescopes, to yield detectable binocular disparities even for very distant targets in our Solar System. In particular, we show that the space telescopes will provide unprecedented opportunities to image the planet Mars, the moons of Jupiter, the rings of Saturn, and other Solar System objects of interest in stereo 3D.  

The HST and JWST missions will overlap significantly: NASA has approved HST for continued operations through at least 2021\footnote{http://www.nasa.gov/press-release/nasa-extends-hubble-space-telescope-science-operations-contract/}, and JWST is projected to launch in October 2018, beginning science operations in mid-2019\footnote{http://jwst.nasa.gov/resources/WebbUpdate\_Winter2016.pdf}.  Thus we anticipate a $\sim$ 2 yr period of overlap in science missions.

HST and JWST both include imaging capability between 0.7 and 1.6 $\mu$m, in several broad and narrow channels, via Wide Field Camera 3 (WFC3) and the Advanced Camera for Surveys (ACS) for HST, and the Near-Infrared Camera (NIRCam) for JWST.  Although JWST has a larger aperture, NIRCam is not diffraction limited at these wavelengths.  However, JWST will achieve comparable or better angular resolution than HST over this range. For the purposes of this work, we adopt the limiting spatial resolution of the pair as that of HST at all common wavelengths.

\section{Technical Requirements}

\subsection{Binocular Disparity}

\begin{figure*}
\centering
\epsscale{1.0}
\includegraphics[scale=0.5]{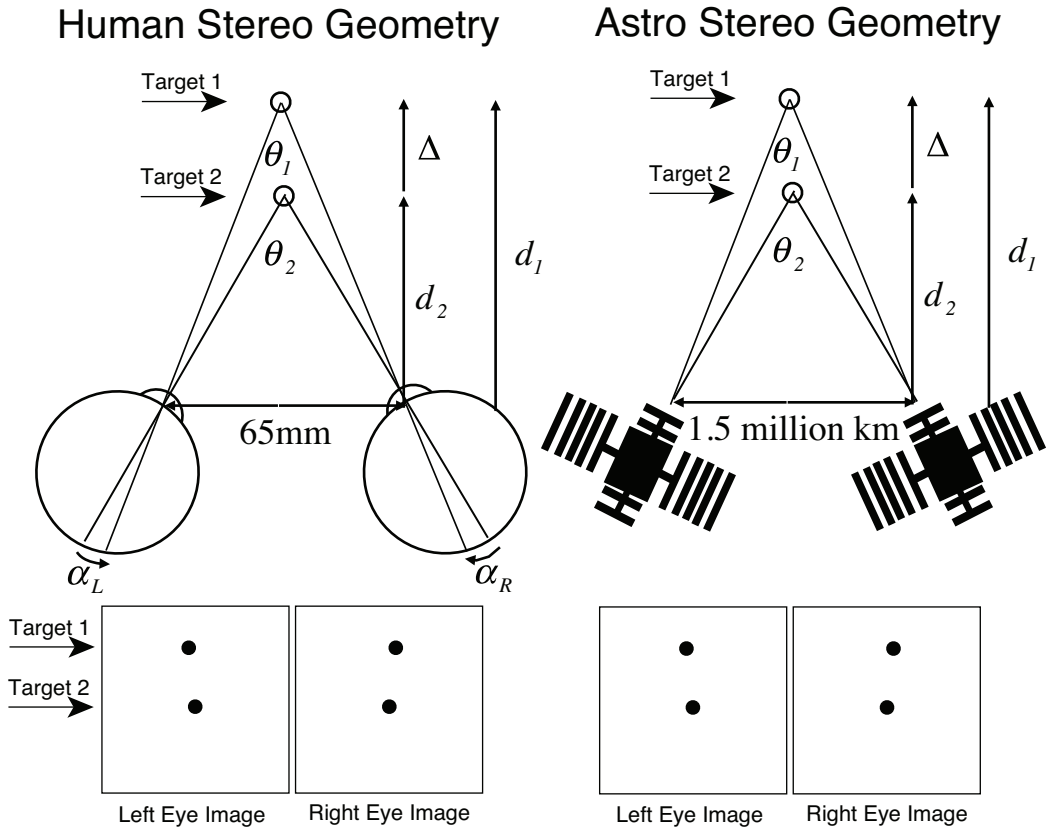}
\figcaption{\label{fig:stereo} {\bf Top Left:} A diagram indicating how human eyes achieve stereoscopic 3D imaging. {\bf Bottom Left:} The difference (disparity) in the resulting left eye and right eye images, which are resolved by the human brain into 3D. {\bf Top and Bottom Right:} The equivalent diagrams for two telescopes with a 1.5 million km separation.}
\end{figure*}

Binocular disparity is defined as the difference between the angles subtended by the lines of sight to two different targets:
\begin{equation}
\delta = \theta_{1} - \theta_{2}
\end{equation}
where $\theta_1$ and $\theta_2$ are the angles subtended by the lines of sight to target points 1 and 2 (Fig. \ref{fig:stereo}). Disparity is equivalently defined by the difference between the angles subtended by the two targets in each eye:
\begin{equation}
\delta = \alpha_R-\alpha_L
\end{equation}
where $\alpha_L$ and $\alpha_R$ are the visual angles subtended by the two targets in the left and the right eyes, respectively. These angles must be resolvable (given the spatial resolution of the sensor) for disparities to be detectable.

To obtain an expression for disparity in terms of distance and depth, we substitute using basic trigonometry:
\begin{equation}
\delta = 2 [tan^{-1}(\frac{I}{2(d_{2}+\Delta)})-tan^{-1}(\frac{I}{2d_{2}})]
\end{equation}
where $\Delta=d_{1}-d_{2}$ is the depth (i.e. distance difference) between one target 1 and target 2, d$_{1}$ is the distance to one target, d$_{2}$ is the distance to the other target, and $I$ is the physical separation between the imaging devices (``eyes'') (Figure \ref{fig:stereo}).  Under the small angle approximation, the equation simplifies to:

\begin{equation}
\delta = \frac{-I\Delta}{d^2_{2}+\Delta d_{2}}
\end{equation}

Human disparity detection thresholds are $\sim$ 1/6 the width of a foveal photoreceptor; foveal photoreceptors subtend approximately 30 arcsec of visual angle. For stereo-astro-photographs to contain detectable disparity signals, the disparities should be at least 1/6 the pixel pitch of the telescopic sensors. The best spatial resolution is 0.07 arcsec for HST/WFC3-UVIS \footnote{http://www.stsci.edu/hst/wfc3/documents/handbooks/currentIHB/c06\_uvis07.html}, and 0.065 arcsec (PSF FWHM) for JWST/NIRCam\footnote{https://jwst.stsci.edu/instrumentation/nircam}, where full spatially sampled ($>$ 2 $\mu$m). In the IR, HST/WFC3-IR is limited to 0.13 arcsec\footnote{http://www.stsci.edu/hst/wfc3/documents/handbooks/currentIHB/c07\_ir07.html}.  If the HST pixel pitch is adopted as the limiting resolution, the minimum detectable disparity will be 0.012 arcsec. Of course, disparities larger than the minimum will produce more compelling effects.

\subsection{Limiting Resolution}

We consider bands in the electromagnetic spectrum  between 0.7 $\mu$m and 1.6 $\mu$m for the purposes of this work.  HST is a 2.4m primary mirror telescope. For the following cases we assume a separation between HST and JWST of 1.5 $\times$ 10$^6$ km, although this distance will vary $\sim$ 10\% every few months due to the planned orbit of JWST $\sim$ 10$^5$ km around the second Lagrange point (L2).

\section{Case Studies for Stereoscopic 3D in our Solar System}

\begin{figure*}
\centering
\includegraphics[scale=0.85]{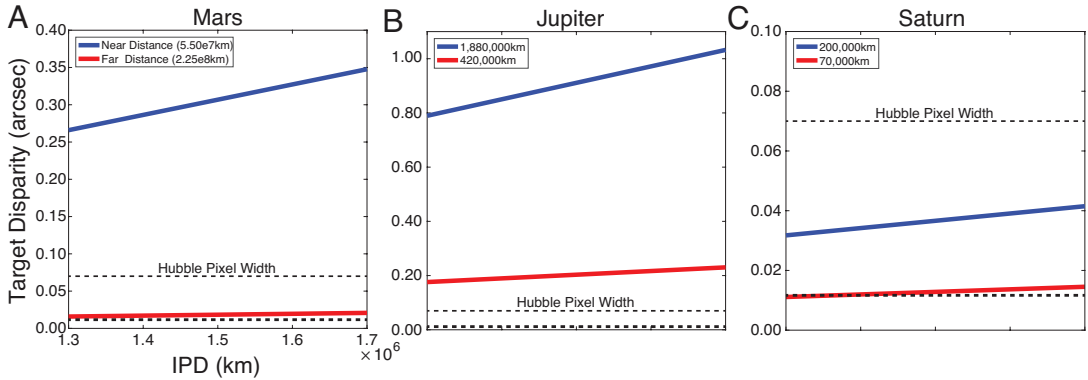}
\figcaption{\label{fig:disparity} Disparities from Mars, the Jupiter moon system, and the Saturnian rings as a function of the physical separation between the telescopes ($\sim$ 1.3-1.7 $\times$ 10$^6$ km).  The top dashed line in each plot represents the Hubble spatial resolution limit; the bottom dashed line represents the disparity detection threshold (1/6 of the resolution). \textbf{A} Disparity between near point and far point  on Mars at the closest (blue) and farthest (red) points in its orbit. \textbf{B} Disparity between the nearest moon (to Earth) of Jupiter and the surface of Jupiter (blue) and the farthest moon of Jupiter and the surface of Jupiter (red) when the planet is 800,000,000 km from the Earth. \textbf{C} Disparity between the outer rings of Saturn and the surface of Saturn (blue) and the inner rings of Saturn and the surface of Saturn (red). The inner rings will yield disparities that are are marginally above the detection threshold.}
\end{figure*}

The disparity and limits for various case studies are shown in Figure \ref{fig:disparity} and discussed in the following sections.

\subsection{Case Study: A Living Mars Globe}

Mars ($d$ $\sim$ 5.5 $\times$ 10$^7$ km) has a radius of 3000km. With a 1.5 million km imaging baseline, when Mars is closest to Earth, the nearest and farthest viewable points on the surface of Mars will yield a disparity of 0.3 arcsec, a large superthreshold disparity given the spatial resolution of the telescopes. The farthest point on Mars' orbit is approximately 5$\times$ farther from Earth than the near point in its orbit. At the far point, the nearest and farthest viewable points on the surface of Mars will yield a disparity of only 0.02 arcsec disparity. This dramatic 25$\times$ change in the magnitude of the  disparity signal is due to the fact that disparity decreases with the square of the distance to the target (Eq. 4). Thus, close to opposition, it is feasible to obtain a stereo-image of Mars, and the spherical 3D shape of Mars will be captured in resolvable stereo 3D. 

The optimal case is slightly less optimistic than above.  First, JWST cannot observe Mars fully at opposition and keep the sunshield oriented properly.  In 2020, Mars is about 0.6 AU (9$\times$10$^7$ km) from JWST when at its closest approach during a visibility window. Second, with NIRCam, saturation can be an issue. Saturation can be avoided by using narrow-band (1\%) filters, the shortest being at 1.64 $\mu$m, but a shorter wavelength filter cannot be used to take a full-disk image. However, it might be possible to do a mosaic using a very small subarray to get an unsaturated full-disk image at 1.4 $\mu$m. In practice, it may not be possible to image under the optimal conditions described above, but it will remain possible to obtain stereoscopic images of Mars with significantly super threshold disparities. 

\subsection{Case Study: Jupiter, Atmosphere, and its Moons}

Jupiter's enormous size (i.e. 140,000 km diameter) and striking atmosphere make it another good candidate for stereoscopic 3D imaging.  Jupiter's moon system provides  an additional opportunity, with an interesting historic tie-in. The Galilean satellites (Io, Europa, Callisto, and Ganymede) were first observed by Galileo in 1610. Galileo was able to infer the rotation of the moons by noting their change in position from night to night. The three-dimensional orbits of the Galilean satellites around Jupiter would be showcased by stereoscopic imagery.  The moons are at distances ranging from 420,000 km to 1,880,000 km from Jupiter ($d=$ 600 million to 1 billion km; average 800 million km). At Jupiter's average 800 million km distance, these moons will yield disparities with respect to the surface that range from 0.2 arcsec to 0.9 arcsec. At the nearest point in Jupiter's orbit, the nearest and farthest viewable points on Jupiter's ``surface'' will yield 0.06 arcsec of disparity.

\subsection{Case Study: Saturn Ring System}

\begin{figure*}
\centering
\includegraphics[scale=0.6]{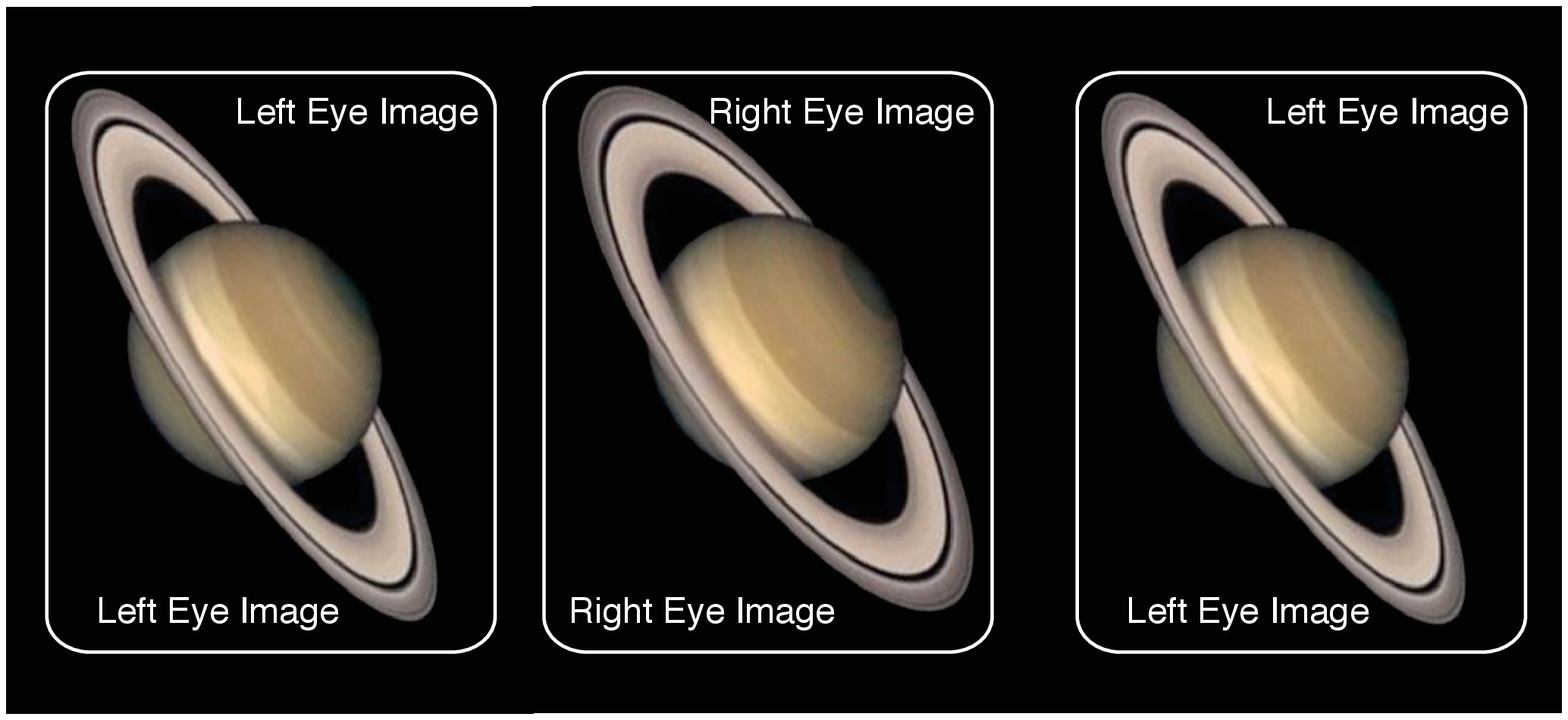}
\figcaption{\label{fig:saturn} Saturn, as viewed by the Hubble Space Telescope during two different seasons in 2004. Atmospheric changes took place during the three month gap between each image capture (e.g. note the differences immediately to the right of the rings in the left- and right-eye images). These local images differences cause perceptual artifacts (e.g. note the shimmery quality of the image when fused). Simultaneous capture eliminates these perceptual artifacts.}
\end{figure*}

Saturn ($d\sim$ 1.3 $\times$ 10$^9$ km) is surrounded by ice particles organized in a ring ranging from 70,000 to 200,000 km from the planet, with a height of $\sim$ 3 km, varying by ring (Figure \ref{fig:saturn}). Additionally, Saturn's rings occasionally show disruption on rapid timescales\footnote{http://www.sci-news.com/space/cassini-image-mini-jet-trail-saturns-f-ring-03946.html} \citep{meinke12}.
To image Saturn's rings in stereoscopic 3D, the  disparity of the 70,000 km rings and the 200,000 km rings from the surface of Saturn.  At 70,000 km, the rings have 0.01 arcsec of disparity (just below our detection threshold), and at 200,000 km the rings have 0.04 arcsec of disparity. Thus, stereo photography of Saturn’s rings may be right at the edge of possible.

\section{Additional Cases}

We note a few additional cases where stereoscopic capabilities can be explored.

\begin{itemize}
\item {\bf Near Earth Asteroid motion.} Although unresolved, the 3D effect can be seen in the asteroid's tumbling path.
\item {\bf Active comet/asteroid.} Several comets have been observed by HST while on their closest approach; combining this track with JWST could produce an intriguing 3D sequence.
\item {\bf Great Red Spot or other cloud features.} It may be possible to discern or monitor topography of cloud tops or other atmospheric features on Jupiter's surface.
\item {\bf Cometary collision/disruption/bending waves/size distribution in Saturn's F ring.} Changes in Saturn's rings due to structural or disruptive events may produce wave effects \citep{esposito08,murray14}.
\end{itemize}

\section{Additional Considerations}

\subsection{Dependence on Telescope Separation}

The 3D effect is significantly stronger when the telescopes are maximally separated, and the targets are as close as possible.  This is a particularly important factor in the proposed Mars observations.

\subsection{Combined Optical and IR Stereoscopic Imaging}

We can achieve this stereoscopic imaging in the regime between 0.7 and 1.6 $\mu$m, which includes a significant number of near-IR bands, with respectively lower resolution and detectable disparity at longer wavelengths.  Given the large change in the appearance of all three targets across this range, it may be interesting to try the effect with multiple bands together.  This would produce a 3D image that could be viewed in multiple wavelengths, which could be layered to reveal relative motions and depths.

\section{Conclusions}

When combined, NASA's Hubble and James Webb Space Telescopes are capable of a unique observation: true stereoscopic 3D imaging, in the optical/near-IR overlap regime.  Separated by 1.5 million kilometers, we can produce 3D images of asteroids, comets, moons and planets in our Solar System, for a variety of astronomical applications. We have outlined the basic  constraints, along with a sampling of case studies.  We anticipate several opportunities during an anticipated 2 yr mission overlap.

\acknowledgements

The authors would like to thank Dr. Jennifer Wiseman, for discussions in which the original idea germinated.

\bibliographystyle{aasjournal}
\bibliography{dissbib}

\begin{thebibliography}{}
\expandafter\ifx\csname natexlab\endcsname\relax\def\natexlab#1{#1}\fi

\bibitem[{{Blakemore}(1970)}]{blakemore70}
{Blakemore}, C. 1970, J. Physiol. (Lond.), 211, 599

\bibitem[{{Burge} \& {Geisler}(2014)}]{burge14}
{Burge}, J., \& {Geisler}, W.~S. 2014, J. Vis, 14

\bibitem[{{Cormack} {et~al.}(1991){Cormack}, {Stevenson}, \&
  {Schor}}]{cormack91}
{Cormack}, L.~K., {Stevenson}, S.~B., \& {Schor}, C.~M. 1991, Vision Research,
  31, 2195

\bibitem[{{Curcio} {et~al.}(1990){Curcio}, {Sloan}, {Kalina}, \&
  {Hendrickson}}]{curcio90}
{Curcio}, C.~A., {Sloan}, K.~R., {Kalina}, R.~E., \& {Hendrickson}, A.~E. 1990,
  J. Comp. Neurol., 292, 497Ð523

\bibitem[{{Esposito} {et~al.}(2008){Esposito}, {Meinke}, {Colwell},
  {Nicholson}, \& {Hedman}}]{esposito08}
{Esposito}, L.~W., {Meinke}, B.~K., {Colwell}, J.~E., {Nicholson}, P.~D., \&
  {Hedman}, M.~M. 2008, \icarus, 194, 278

\bibitem[{{Meinke} {et~al.}(2012){Meinke}, {Esposito}, {Albers}, \& {Srem{\v
  c}evi{\'c}}}]{meinke12}
{Meinke}, B.~K., {Esposito}, L.~W., {Albers}, N., \& {Srem{\v c}evi{\'c}}, M.
  2012, \icarus, 218, 545

\bibitem[{{Murray} {et~al.}(2014){Murray}, {Cooper}, {Williams}, {Attree}, \&
  {Boyer}}]{murray14}
{Murray}, C.~D., {Cooper}, N.~J., {Williams}, G.~A., {Attree}, N.~O., \&
  {Boyer}, J.~S. 2014, \icarus, 236, 165

\end{thebibliography}
\include{dissbib}
\end{document}